\documentclass[aps,pra,showpacs,groupedaddress,floatfix,nofootinbib]{revtex4}
\usepackage{amsmath}

\usepackage{graphicx}

\usepackage{color}

\begin{document}

\title{Comment on: ``Bound states and the potential parameter spectrum''. J. Math.
Phys. \textbf{67}, 062103 (2020)}

\author{Francisco M. Fern\'andez}\email{fernande@quimica.unlp.edu.ar}

\affiliation{INIFTA, Blvd. 113 y 64 S/N,
 Sucursal 4, Casilla de Correo 16, 1900 La Plata, Argentina}

\begin{abstract}
We analyze the application of the ``tridiagonal representation
approach'' (TRA) to the Schr\"{o}dinger equation for some simple,
exactly-solvable, quantum-mechanical models. In the case of the
Kratzer-Fues potential the mathematical reasoning appears to
exhibit a serious flaw that invalidates the result and the
expression for the energy does not appear to be correct. We also
show that the well known Frobenius method, which resembles the
TRA, is far simpler, clearer and more elegant; in addition to give
the correct result.
\end{abstract}


\maketitle

In a recent paper Alhaidari and Bahlouli\cite{AB20} (AB from now
on) applied the so called tridiagonal representation approach
(TRA) and the concept of potential parameter spectrum (PPS) to the
nonrelativistic Schr\"{o}dinger equation. In this way they
obtained the spectrum of some well known exactly-solvable models
like the Kratzer-Fues potential, the Morse oscillator and the
Rosen-Morse potential. In what follows we analyze their results
using the Kratzer-Fues potential as illustrative example and
compare the TRA PPS results with those coming from the Frobenius
method\cite{BO78}.

AB stated that they resorted to atomic units $\hbar =m=1$ and
wrote the radial operator as
\begin{equation}
H=-\frac{1}{2}\frac{d^{2}}{dr^{2}}+\frac{l(l+1)}{2r^{2}}+V(r),
\label{eq:H}
\end{equation}
where $l=0,1,\ldots $ is the angular momentum quantum number. They
defined the operator $J=H-EI$ where $I$ is the identity operator.
It is worth mentioning that in atomic units only the electron mass
is set to unity\cite {F20} which is not the case here as discussed
below.

In order to apply the TRA AB chose the basis set o functions
\begin{equation}
\phi _{n}(r)=\left( \lambda r\right) ^{\mu }e^{-\lambda
r/2}L_{n}^{\nu }(\lambda r),  \label{eq:basis}
\end{equation}
``where $L_{n}^{\nu }(\lambda r)$ is a Laguerre polynomial,
$\lambda $ is a positive scale parameter having dimension of
inverse length, and $\{\mu ,\nu \}$ are dimensionless real
parameters such that $\nu >-1$.'' First, notice that if they were
using atomic units then $r$ would be dimensionless\cite {F20} and
$\lambda r$ would have units of length$^{-1}$ so that it could not
be a suitable argument of the Laguerre polynomials. Choosing
$\hbar =m=1$ is
equivalent to multiplying the Schr\"{o}dinger equation $\mathcal{H}\psi =%
\mathcal{E}\psi $, $\mathcal{H}=\mathcal{T}+\mathcal{V}$, by
$m/\hbar ^{2}$
and defining $T=m\mathcal{T}/\hbar ^{2}$, $V=m\mathcal{V}/\hbar ^{2}$ and $%
E=m\mathcal{E}/\hbar ^{2}$. Note that $T$, $V$ and $E$ have units of length$%
^{-2}$ which is consistent with all the equations and results
derived by AB. If they were in fact using atomic units (as they
claimed) all their equations and results would be
dimensionless\cite{F20}. Second, if $\mu $ is negative the basis
functions are singular at origin; therefore, $\mu $ should be
positive. The use of the kinetic-energy operator as in Eq.~(\ref
{eq:H}) requires that the wavefunction vanish at origin which forces $\mu >0$%
.

The authors applied $J$ to $\phi _{n}$ and obtained
\begin{eqnarray}
-\frac{2}{\lambda ^{2}}J\phi _{n}(r) &=&x^{\mu }e^{-x/2}\left[ \frac{d^{2}}{%
dx^{2}}+\left( \frac{2\mu }{x}-1\right) \frac{d}{dx}+\frac{\mu
(\mu
-1)-l(l+1)}{x^{2}}\right.  \nonumber \\
&&\left. -\frac{\mu }{x}+\frac{1}{4}-\frac{2V(x)}{\lambda ^{2}}+\frac{2E}{%
\lambda ^{2}}\right] L_{n}^{\nu }(\lambda r),
\end{eqnarray}
where $x=\lambda r$. It seems that the authors simply wrote $%
V(x)=V(r)=V(x/\lambda )$ in this equation. Then, they resorted to
the differential equation for the Laguerre polynomials and other
well known properties of these functions in order to obtain a
suitable equation for the TRA. To this end they required that
$\lambda ^{2}=-8E$ (note that $E$ have units of length$^{-2}$ as
conjectured above) and
\begin{equation}
V(x)=\frac{\lambda ^{2}}{2}\frac{\alpha x+\beta
}{x^{2}}=\frac{\alpha \lambda }{2r}+\frac{\beta }{r^{2}},
\label{eq:V(x)_AB}
\end{equation}
``where $\alpha $ and $\beta $ are arbitrary dimensionless
parameters''.
First, note that this Kratzer-like potential depends on the energy through $%
\lambda $ unless one removes such dependence by a suitable choice
of $\alpha $. Second, $\alpha $ and $\beta $ cannot be arbitrary;
in fact, taking into account that $\lambda >0$ then $\alpha $
should be negative in order to have
bound states. Besides, a Kratzer-Fues oscillator requires that $\beta >0$%
\cite{K20,F26}. AB stated that ``the electric charge is $Z=\alpha \lambda /2$%
'' which is most curious because $Z$ should be negative in order
to have
bound states as argued above (note that the resulting potential is $V(x)=%
\frac{Z}{r}+\frac{\beta }{r^{2}}$). Besides, the Kratzer-Fues
potential was proposed as a phenomenological model for the
analysis of the vibration-rotation spectra of diatomic
molecules\cite{K20,F26} and, consequently, the negative
coefficient of $1/r$ is not expected to be an electric charge. It
is also worth mentioning that if $\alpha $ is dimensionless then
the \textit{electric charge} $Z$ is given in units of
length$^{-1}$.

After a tedious analysis of the equations that included comparison
of their recursion relation with the one for the dual Hahn
polynomial that AB showed
in an appendix, followed by the analysis of the spectrum of a polynomial $%
S_{n}(z^{2};a,b,c)$ they concluded that the asymptotics of this
polynomial is sinusoidal with an amplitude $A(z)$ that is required
to vanish. In this way AB arrived at the PPS formula
\begin{equation}
\beta =\left( k+l+1+\frac{2Z}{\lambda }\right) \left( k-l+\frac{2Z}{\lambda }%
\right) ,  \label{eq:beta_PPS}
\end{equation}
``where $k=0,1,\ldots ,N$ and $N$ is the integer less than or equal to $%
-\left( \frac{1}{2}+\frac{2Z}{\lambda }\right) $.'' It is clear
that $0 \leq N$ $\leq -\left( \frac{1}{2}+\frac{2Z}{\lambda
}\right) $ makes sense only if $Z=\alpha \lambda /2\leq -\lambda
/4$ as argued above.

From one of the two roots of the PPS formula (\ref{eq:beta_PPS})
they derived the following equation for the energy:
\begin{equation}
E_{k}=-\frac{Z^{2}}{2\left[ k+\frac{1}{2}+\sqrt{\beta +\left( l+\frac{1}{2}%
\right) ^{2}}\right] ^{2}}.  \label{eq:E_k_AB}
\end{equation}
This expression resembles the well known spectrum of the
Kratzer-Fues potential, except for the slight difficulty that the
quantum number $k$ is bounded by $N$ as discussed above. However,
AB overcame this problem in a
most striking way; they stated that ``Note that the condition $N\leq $ $%
-\left( \frac{1}{2}+\frac{2Z}{\lambda }\right) $ allows the
non-negative integer $k$ to assume all values from zero to
infinity''. There is no doubt that this argument is false and the
AB's lengthy and tedious derivation of the eigenvalues of the
Kratzer-Fues potential appears to have a flaw. It is also worth
mentioning that if we multiply $E_{k}$ by the atomic unit of
energy\cite{F20} then the resulting spectrum will have the wrong
units of energy$\times $length$^{-2}$. On the other hand, note
that $\hbar ^{2}Z^{2}/m $ exhibits the expected units of energy.

It is not our purpose to put forward a correct proof of AB's
approach because in our opinion the TRA PPS is far too cumbersome
and intricate to deserve further attention. Instead, we will solve
the Schr\"{o}dinger equation with the Kratzer-Fues potential by
means of the well known Frobenius method\cite{BO78}. The reason
for this choice is that the Frobenius method resembles, in some
ways, the TRA\ PPS and at the same time is far simpler, clearer
and more elegant.

To begin with we write the Kratzer-Fues potential as
\begin{equation}
V(r)=-\frac{A}{r}+\frac{B}{r^{2}},  \label{eq:V(r)_KF}
\end{equation}
where $A>0$ and $B>0$ are dimensionless parameters\cite{F20}. This
interaction is repulsive at short interatomic distances and
attractive at large ones, as expected for a diatomic
potential\cite{K20,F26}. In order to solve the radial eigenvalue
equation we choose the ansatz
\begin{equation}
\varphi (r)=r^{s}e^{-\alpha r}u(r),  \label{eq:ansatz}
\end{equation}
where $s>0$ and $\alpha >0$ in order to have the correct
asymptotic behavior at $r=0$ and $r\rightarrow \infty $,
respectively. The function $u(r)$ satisfies the differential
equation

\begin{eqnarray}
&&-\frac{1}{2}u^{\prime \prime }(r)+\left( \alpha
-\frac{s}{r}\right)
u\,^{\prime }(r)+\frac{\left( \alpha s-A\right) }{r}u{\left( r\right) }+%
\frac{\left[ 2B+l(l+1)-s\left( s-1\right) \right] u{\left( r\right) }}{2r^{2}%
}  \nonumber \\
&&-\frac{\left( \alpha ^{2}+2E\right) u\left( r\right) }{2}=0.
\label{eq:dif_u_1}
\end{eqnarray}
We choose $s$ and $\alpha $ so that the last two terms vanish
\begin{equation}
s=\frac{\sqrt{8B+4l^{2}+4l+1}+1}{2},\;\alpha =\sqrt{-2E},
\label{eq:s,E}
\end{equation}
and the differential equation reduces to
\begin{equation}
-\frac{1}{2}u^{\prime \prime }(r)\,+\left( \alpha
-\frac{s}{r}\right) u\,^{\prime }(r)+\frac{\left( \alpha
s-A\right) }{r}u{\left( r\right) =0.} \label{eq:dif_u_2}
\end{equation}
The first expression in Eq.~(\ref{eq:s,E}) comes from the indicial
equation. Note that the solution is regular at origin provided
that $B\geq -(2l+1)^{2}/8\geq -1/8$ and compare it with AB's
result $\beta >-\left( l+1/2\right) ^{2}$. We will come back to
this discrepancy later on.

If we insert the power-series expansion
\begin{equation}
u(r)=\sum_{j=0}^{\infty }c_{j}r^{j},  \label{eq:u_series}
\end{equation}
we find that the expansion coefficients $c_{j}$ satisfy the
recurrence relation
\begin{equation}
c_{{j+1}}=2\frac{\alpha \left( j+s\right) -A}{\left( j+1\right)
\left( j+2s\right) }c_{{j}},\;j=0,1,\ldots ,\;c_{0}\neq 0.
\label{eq:rec_rel}
\end{equation}
It follows from
\begin{equation}
\lim\limits_{j\rightarrow \infty }j\frac{c_{j+1}}{c_{j}}=2\alpha ,
\label{eq:lim_cj}
\end{equation}
that the series (\ref{eq:u_series}) converges for all $r$ and that
$u(r)$ behaves as $e^{2\alpha r}$ as $r\rightarrow \infty $.
Therefore, $\varphi (r) $ is not square integrable unless $\alpha
=A/(n+s)$, $n=0,1,\ldots $. We thus obtain the allowed energies
\begin{equation}
E_{n,l}=-\frac{A^{2}}{2\left( n+s\right) ^{2}}=-\frac{A^{2}}{2\left[ \sqrt{%
2B+\left( l+1/2\right) ^{2}}+n+1/2\right] ^{2}}.  \label{eq:E_n,l}
\end{equation}
From this expression we realize that equation (\ref{eq:E_k_AB}) is
incorrect (unless $2\beta $ is substituted for $\beta $). This
discrepancy is consistent with the one mentioned previously.

Summarizing: We have shown that the application of the TRA PPS to
the Kratzer-Fues potential\cite{AB20} has a serious flaw with
regard to the values of the radial quantum number $k$. The
resulting expression for the spectrum does not appear to be
correct. In addition to it, the well known Frobenius
method\cite{BO78}, which resembles the TRA\ PPS and is taught in
undergraduate courses on quantum mechanics and quantum
chemistry\cite{P68}, is far simpler, clearer and more elegant.

\end{document}